\documentclass{article}
\newcommand{\bfr}{\begin{flushright}}
\newcommand{\efr}{\end{flushright}}
 
\begin{document}
\title{Aharonov-Bohm scattering by vortices of dimensionally-reduced
 Yang-Mills field
}
\author{Kiyoshi Shiraishi\\
Department of Physics, Faculty of Science, Ochanomizu University,\\
1-1 Otsuka 2, Bunkyou-ku, Tokyo 112, Japan\\ 
and\\
Atsushi Nakamula\\
Department of Physics, Tokyo Metropolitan University,\\
Setagaya-ku, Tokyo 158, Japan
}
\date{Czechoslovak Journal of Physics
Volume 42,\\ Number 3 (1992), 285-289
}
\maketitle
\begin{abstract}
If two dimensions of six-dimensional space-time are compactified, a
topological configuration of Yang-Mills gauge field appears as a cosmic
string in four dimensions, whose thickness is of the same order as the
size of the compact space. We consider scattering of low-energy
fermions by this object. 
\end{abstract}

\bigskip

Recently, the present authors and Hirenzaki have verified the
existence of a vortex-like classical solution in a six-dimensional
Yang-Mills (YM) system with partially compactified space \cite{1}. This
object appears to be a cosmic string (CS) \cite{2} in our space-time. It
is notable that this string arises without the need for elementary
scalar fields. CSs of this type might appear in Kaluza-Klein models
including the anomaly-free supergravity model proposed in
ref.~\cite{3}. This CS may be connected with scenarios of making
clusters of galaxies in (multidimensional) cosmology.

In ordinary CS scenarios of formation of a large scale structure, it
is necessary to consider gravitational interactions of cosmic strings.
Their characteristics have been understood by analytical and/or
numerical methods. Interactions of matter with CS without gravity have
also been investigated in various models, which were principally
related to topics of proton decay in unified theories \cite{4,5}.

The most peculiar interaction independent of details in models is known
as the Aharonov-Bohm (AB) scattering \cite{6}. The scattering of
particles occur in the region far away from the axis of CS, where the
gauge configuration is expressed as pure-gauge. This effect is due to
phases of wave functions of the particles. 

In this letter, we consider
the AB scattering in the CS model of our six-dimensional
system \cite{1}. The main distinction from the ordinary four-dimensional
models mentioned above is that matter fields also come from
six-dimensional matter fields. We must be careful of the fact that a
particle with a unit charge is not scattered by a CS with a quantized
flux (in the absence of gravity). This phenomenon is understood as the
same situation of the ``string'' in the Dirac monopole \cite{7}.
For the simplest case, namely, in a $U(1)$-Higgs system, there is no
scattering in the case with only a unit gauge coupling constant.

Since our model consists of YM field, no charged field coupled in
arbitrary strength is allowed. In this paper we introduce fermions in a
fundamental representation of $SU(2)$ as a matter field. As we will see
later, the effective charge of this fermion is half of the unit when
the space-time is reduced.

Now, at the beginning, we review the CS-like solution to
six-dimensional YM equations in the case of a constant geometry.

For a constant background space-time geometry, we choose a sphere as an
extra compactified two-space. Further we give the parametrization of
the components on this extra space as follows,
\begin{eqnarray}
& &A_\theta=\frac{1}{2}\left(\begin{array}{cc}
0 & -i\Phi e^{-i\phi}\\
i\Phi^* e^{i\phi} & 0
\end{array}
\right)\,,\label{eq1}\\
& &A_\phi=-\frac{1}{2}\sin\theta\left(\begin{array}{cc}
0 & \Phi e^{-i\phi}\\
\Phi^* e^{i\phi} & 0
\end{array}
\right)+\frac{1}{2}\left(\begin{array}{cc}
1-\cos\theta & 0\\
0 & -(1-\cos\theta)
\end{array}
\right)\,,\label{eq2}
\end{eqnarray}
where $\theta$ and $\phi$ are the standard spherical coordinates and
$\Phi$ is assumed to depend only on the four-dimensional coordinates.
Besides these components, we take out the following ``$U(1)$'' part as
the four-dimensional components of gauge field
\begin{equation}
A_\mu=A_\mu\times\frac{1}{2}\left(\begin{array}{cc}
1 & 0\\
0 & -1
\end{array}
\right)\,,
\label{eq3}
\end{equation}

By these parametrizations, the YM equations can be shown to be reduced
to those of the Abelian-Higgs model in which $A_\mu$ and $\Phi$ are
regarded as the $U(1)$ gauge field and the Higgs field, respectively
\cite{1}. In these equations the effective Higgs self-coupling is
related with the gauge coupling constant. 

Moreover the relation between
the $U(1)$ gauge coupling and the Higgs coupling constant causes that a
vortex solution is given as a self-dual solution to first-order
simultaneous differential equations. Of course, this vortex solution is
self-dual configuration of YM field in the original form. Namely, the
classical solution represents embedding of an instanton into the
four-space which consists of extra space plus the $x$-$y$ plane. The
width of this string is of the same order of size as the extra space.

The classical solution with $n$-vorticity centered at the origin has
the following behaviour:
\begin{eqnarray}
& &\Phi=0\quad\mbox{at the origin,}\quad\Phi=e^{in\varphi}\quad
\mbox{at the spatial infinity,}\label{eq4}\\
& &A_\varphi=0\quad\mbox{at the origin,}\quad
A_\varphi=-n\quad
\mbox{at the spatial infinity,}\label{eq5}
\end{eqnarray}
where $r$ and $\varphi$ are the polar coordinates of $x$-$y$ plane. In
this description we do not use the orthonormal coordinate basis but a
general coordinate basis. If $r$ is infinity, the field strengths of
$SU(2)$ are zero, i.e., expressed in a pure-gauge form. The magnitude of
flux is determined by the winding number, $n$. 

It is important to note
that the flux of the string is not arbitrary as in usual cosmic string
models.

Now, we consider a fermion in the fundamental representation of $SU(2)$
as a matter field. In six dimensions, a Dirac fermion has at least
eight complex components. Here, for simplicity, we deal with a massless
Weyl fermion, while an extension to a general case is straightforward.
The Dirac matrices in six dimensions are defined as follows;
\begin{eqnarray}
& &\Gamma^\mu=\gamma^\mu\otimes\sigma^1\,,\quad(\mu=0, 1, 2,
3)\,,\label{eq6}\\ & &\Gamma^4=\gamma_5\otimes\sigma^1\,,\label{eq7}\\
& &\Gamma^5=1_4\otimes\sigma^2\,,\label{eq8}
\end{eqnarray}
where $\gamma^\mu$, $\gamma_5$ and $1_4$ are the four-dimensional Dirac
matrices, chiral and unit matrices, respectively. The Weyl condition
is $\Gamma_7\Psi=\gamma_5\otimes\sigma^3\Psi=+\Psi$.

When the compact space has non-zero curvature, the fermions acquire
super heavy masses in four dimensions. This is also true when the
fermions are coupled only with the gauge field which is expressed
merely by pure gauge. In the present model, we can have massless
particle states by introducing a coupling with a topologically
non-trivial gauge field configuration \cite{8}. In ref.~\cite{1}, we
have introduced an extra $U(1)$ gauge field to stabilize the compact
space. Here we make use of it to obtain massless modes of fermion.

The Dirac equation is written as follows,
\begin{equation}
\Gamma^MD_M\Psi=\Gamma^M\left(\partial_M+\frac{1}{4}\omega_{MAB}\Gamma^{AB}
+iA_M+iB_M\right)\Psi=0\,,
\label{eq9}
\end{equation}
where $\Gamma^{AB}=\frac{1}{2}[\Gamma^A, \Gamma^B]$
and
$B_M$ is the newly introduced
$U(1)$ gauge field which has the following monopole-like classical
configuration 
\begin{equation}
B_M=\frac{1}{2}(\cos\theta\pm 1)\,.\label{eq10}
\end{equation}

We consider eigenstates of four-dimensional chiral matrix as
$\gamma_5\psi=\psi$. We assume the string is sufficiently narrow in
comparison with the wavelength of the matter field. We seek for the
lowest energy states in (outside) vacuum, i.e., the zero modes of the
covariant derivative which acts in the extra space. Other fermionic
modes besides the zero mode have masses of the order of an inverse of
the scale of the extra space, in our world. 

There exist two of such
zero-mode eigenfunctions which correspond to two components of the
fundamental representation.
Those are the following ones, 
\begin{eqnarray}
\Psi_I(\theta, \phi, \varphi)&=&
\left(
\begin{array}{c}
\cos\frac{1}{2}\theta\, e^{i\phi/2} \\
e^{-in\varphi}\sin\frac{1}{2}\theta\, e^{3i\phi/2}
\end{array}
\right)\,,
\label{eq11}\\
\Psi_{II}(\theta, \phi, \varphi)&=&\left(
\begin{array}{c}
-\sin\frac{1}{2}\theta\, e^{-i\phi/2} \\
e^{-in\varphi}\cos\frac{1}{2}\theta\, e^{i\phi/2}
\end{array}
\right)\,,
\label{eq12}
\end{eqnarray}

In these descriptions, the column denotes the components of $SU(2)$
doublet and we have omitted to write the eigenfunction $\psi$ of
four-dimensional chiral matrix, i.e., $\gamma_5\psi=\psi$.

It is an essential difference from the eigenfunctions of ordinary
Kaluza-Klein mode function that they depend on the azimuthal coordinate
$\varphi$. 

From now on, we can trace the same way as in ref.~\cite{5}.
The whole eigenfunctions are given as direct products of what was
derived above and of the following wave function
\begin{equation}
P_{kl}^{(\pm)}(r, \varphi)=a_k\left(
\begin{array}{c}
J_{\pm(l+n/2)}(kr) \\
\pm i J_{\pm(l+n/2+1)}(kr) e^{i\varphi}
\end{array}
\right) e^{il\varphi-i\omega t}\,,
\label{eq13}
\end{equation}
where $J_n(x)$'s are the Bessel functions, $k$ is the energy of the
partial wave and $l$ is the label of the angular momentum. The two
components here are those of a Weyl fermion.

In the approximation of infinitesimal string thickness, we can get the
differential cross section by the standard way \cite{5} with the
assumption that the functional form of the extra-dimensional portion of
incoming and outgoing waves are the same. The cross section per unit
length (along the string) is,
\begin{equation}
\frac{d\sigma}{d\Theta}=\frac{\frac{1}{2}[1-(-1)^n]}{2\pi
k\sin^2\frac{\Theta}{2}}\,,
\label{eq14}
\end{equation}
where $\Theta$ is a scattering angle.

As a conclusion of this, we can observe scattering of a massless
fermion in the fundamental representation of $SU(2)$ by the string which
has the flux with any odd-numbered unit, in our model.

We must emphasize that in our model the charge of the fermions differs
by a factor $\frac{1}{2}$ from the usual one, giving a scattering cross
section with maximal Aharonov-Bohm enhancement. The cause of this is
that the coupling constant between $A_\mu$ and $\Psi$ is half of that
between $A_\mu$ and
$\Phi$. The factor $\frac{1}{2}$ can be seen in equation (\ref{eq3}). 

This model
will be extended to have a greater group and a higher-dimensional extra
space. In such a case it is expected that there occurs the AB
scattering by the CS of the present kind. Obviously, matter fields in
the faithful representations of the group, such as the adjoint
representation, are never scattered by the CS.

In the case that the wavelength is comparable with the string
thickness, namely when the thickness cannot be negligible, we should
treat the wave function inside the string core exactly. Then we analyze
whether mixing among Kaluza-Klein excitation states occurs or not.
Since dependence in extra-dimensional coordinates of eigenfunctions
varies with the four-dimensional coordinate $r$, it is suitable to study
the effects of extra space in a very early, multi-dimensional,
universe. Further, it is a challenging problem for numerical analysis.
This will be a future subject.

\bigskip

\bigskip

\bigskip

The authors would like to thank S. Hirenzaki for some useful comments.
One of the authors (KS) would like to thank A. Sugamoto for reading
this manuscript. KS is indebted to Soryuusi shogakukai for financial
support. He also would like to acknowledge the financial aid of Iwanami
F\=ujukai.



\begin{thebibliography}{99}
\bibitem{1} Nakamula A., Hirenzaki S., Shiraishi K.: Nucl. Phys. B339
(1990) 533.

Nakamula A., Shiraishi K.: Mod. Phys. Lett. A5 (1990)
1109.

Nakamula A., Shiraishi K.: Prog. Theor. Phys. 84
(1990) 1100.

\bibitem{2} Vilenkin A.: Phys. Rep.121 (1985) 263.
\bibitem{3} Randjbar-Daemi S., Salam A., Sezgin E., Strathdee J.:
Phys. Lett. B151 (1985) 351.

\bibitem{4} Brandenberger R. H., Perivolaropoulos L.: Phys. Lett.
B205 (1988) 396.

Brandenberger R. H., Davis A. C., Matheson A. M.:
Nucl. Phys. B307 (1988) 909; Phys. Lett. B218 (1989) 304.

Davis A. C., Perkins W.: Phys. Lett. B218 (1989) 37.

Gregory R., Davis A. C., Brandenberger R. H.: Nucl.
Phys. B323 (1989) 187.

Perivolaropoulos L., Matheson A. M., Davis A. C.,
Brandenberger R. H.: Phys. Lett. B245 (1990) 556.

Matheson A. M., Perivolaropoulos L., Perkins W., Davis A.
C., Brandenberger R. H.: Phys. Lett. B248 (1990) 263.

Alford M. G., March-Russell J., Wilczek F.: Nucl. Phys.
B328 (1989) 140.

Kogan Ya. I., Selivanov K. G.: Zh. Eksp. Teor. Fiz. 97
(1990) 387.

de Sousa Gerbert Ph.: Phys. Rev. D40 (1989) 1346.
\bibitem{5} Alford M. G., Wilczek F.: Phys. Rev. Lett. 62 (1989) 1071.
\bibitem{6} Aharonov Y., Bohm D.: Phys. Rev. 115 (1959) 485.
\bibitem{7} Dirac P. A. M.: Proc. R. Soc. London A133 (1931) 60.
\bibitem{8} Randjbar-Daemi S., Salam A., Strathdee J.: Nucl. Phys.
B214 (1983) 491.

\end{thebibliography}
\end{document}